# Reality and super-reality: properties of a mathematical multiverse

Alan McKenzie

*Alumnus of the University of St Andrews*
*School of Physics and Astronomy, North Haugh, St Andrews, Scotland KY16 9SS*

**Abstract**

Ever since its foundations were laid nearly a century ago, quantum theory has provoked questions about the very nature of reality. We address these questions by considering the universe – and the multiverse – fundamentally as complex patterns, or mathematical structures. Basic mathematical structures can be expressed more simply in terms of emergent parameters. Even simple mathematical structures can interact within their own structural environment, in a rudimentary form of self-awareness, which suggests a definition of reality in a mathematical structure as simply the complete structure. The absolute randomness of quantum outcomes is most satisfactorily explained by a multiverse of discrete, parallel universes. Some of these have to be identical to each other, but that introduces a dilemma, because each mathematical structure must be unique. The resolution is that the parallel universes must be embedded within a mathematical structure – the multiverse – which allows universes to be identical within themselves, but nevertheless distinct, as determined by their position in the structure. The multiverse needs more emergent parameters than our universe and so it can be considered to be a superstructure. Correspondingly, its reality can be called a super-reality. While every universe in the multiverse is part of the super-reality, the complete super-reality is forever beyond the horizon of any of its component universes.

## 1. A first approach to reality

Philosophers have been arguing about reality and its interpretation since even before Plato and Aristotle. Physicists came late to the party, but they brought with them a heady brew – the quantum theory. One of its founders, Werner Heisenberg, wrote in 1958 that

> "the idea of an objective real world whose smallest parts exist objectively in the same sense as stones or trees exist, independently of whether or not we observe them...is impossible."

Quantum mechanics, which made predictions that could be physically tested, has even been interpreted by some as saying that reality depends upon human consciousness:

> "The doctrine that the world is made up of objects whose existence is independent of human consciousness turns out to be in conflict with quantum mechanics and with facts established by experiment"
>
> d'Espagnat (1979)

In this paper, we are going to take the opposite view from physicist and philosopher Bernard d'Espagnat: we shall claim that self-awareness is a product of an independent reality rather than the other way around. In support of our position we shall describe heuristically how this might come about in Section 3.

In the meantime, to get straight to the point without being diverted by ontological and epistemological discussions (for those, a good starting point might be Leifer 2014), let us make a first attempt at a working definition of reality:

Reality (I)

*Reality is the complete set of quantum fields extending throughout the whole of spacetime that comprises our block universe.*

Before going further, we need to unpack this definition to see if it works. The block universe may be thought of as the four-dimensional block of spacetime that encapsulates the complete past and future history of our universe. (The notion of the block universe, of course, has a respectable provenance, and, indeed, the block universe may even be demonstrated experimentally (McKenzie 2016(b)).) So, our definition is saying that reality is the complete past and future history of all of the quantum fields that go to make up our universe. That must include all of the particles that arise from stable excitations of the appropriate quantum fields as well as all of their interactions.

Symbolically, the dependence of our universe $U$ on quantum fields is

$$U = \{\varphi_i : 1 \leq i \leq i_{max}\} \qquad (1)$$

where $i$ denotes the $i^{th}$ quantum field, $\varphi_i$, $i_{max}$ is the total number of quantum fields and the curly brackets represent the set.

A consequence of our definition of reality is that it cannot be expressed more fundamentally in terms of contemporary physics. For instance, if we were instead to define reality in terms of just the fundamental particles of the Standard Model, then that definition would be open to the charge that the particles are themselves describable as excitations of the appropriate quantum fields. On the other hand, if we try to extend our definition more deeply than the quantum fields, then all we are left with, like the grin on the Cheshire cat, is the essential pattern of the quantum fields. There are straws in the wind that the patterns of quantum fields and their interactions may be further reducible into yet more fundamental patterns (see, for instance, Arkani-Hamed et al. 2014) but there is no hint of any return to *stuff* – it's patterns all the way down!

## 2. Philosophical aspects of accepting a mathematical stratum as the fundamental level of reality

Patterns, of course, are just mathematical structures, but, at least for some in the physics community, it seems to be less provocative to claim that the universe is, fundamentally, a pattern rather than to say that it is a mathematical structure.

It may be easier to accept that that the fundamental level of our reality is a mathematical structure by asking how, in a Theory of Everything, we might explain our universe, particularly including space and time itself. Clearly, our explanation cannot depend upon any property of the universe that involves space or time, since that is what we are trying to explain. A pure mathematical structure satisfies this requirement. This is well illustrated, for example, in papers by physicists who attempt to model the emergence of spacetime: their approach is to start with an abstract (quantum) mathematical structure in Hilbert space and thence demonstrate the emergence of gravity. Since gravity is a property of the (general relativistic) structure of spacetime, they are effectively demonstrating the emergence of spacetime itself (see, for example, ChunJun, Carroll, and Michalakis (2017) or Raasakka (2017)).

This will be one of the claims of our paper, that our reality is purely a mathematical structure. To that extent, at least, we are in agreement with Tegmark (2008). We shall see at the end of Section 5 that such a philosophy, which is, in essence, one of ontic structural realism (Berghofer 2018), makes it easier to accept the idea of parallel universes, because a pattern (a universe) may be duplicated without controversy.

In spite of the above, many people harbour the suspicion that, no matter how sophisticated a mathematical structure we may conjecture, that structure can never be more than a mere *description* of reality, rather than reality itself. For these people, we have not breathed fire into the equations. They find it hard to accept that their own self-awareness can be just a mathematical structure. So goes their argument, along lines reminiscent of Dr Johnson's refutation of Bishop Berkeley's immaterialist philosophy by kicking a stone and feeling the reverberation in his body.

The next section is essentially a response to that argument as well as to that of d'Espagnat, who, as we saw in Section 1, did not accept the independence of reality from human consciousness. We shall show that a simple mathematical structure can, in principle, demonstrate an awareness of itself. The structure that we use will also serve later in our discussion of super-reality in Appendix A1.

## 3. How a simple mathematical structure can in principle be self-aware

It is easy to see why many physicists reject such a statement. While they may accept that our universe can be *described* by mathematics, it is seemingly a step too far to think that it can *be* mathematics. Nevertheless, a case can be argued to make that idea more palatable by considering a version of a two-dimensional cellular automaton, Conway's Game of Life (Gardner 1970), in which, as we shall see, there can be a structure that is "aware" of its environment. (It is sometimes helpful to use such automata rather than to appeal to our own subjective experience as self-aware beings in our own universe, with all of the associated baggage that such experience would entail.)

The main attraction of the Game of Life is that its very simple set of rules can lead to solutions of interesting complexity. The game evolves within a two-dimensional ($p$, $q$) matrix, the cells of which live or die, and it is typically played out on a computer monitor screen which displays successive generations (labelled $h$) of the matrix. The pattern in any generation (generation $h$) is transformed according to the rules, and the resulting pattern is then displayed as the next ($h$ + 1) generation. Pragmatically, the matrix is finite in extent which means that boundary conditions (for instance, toroidal) must be chosen.

The rules are that (1) a live cell with either two or three live neighbours (formally called the Moore neighbourhood) will live on to the next generation but will die otherwise and (2) a dead cell with exactly three live neighbours will become alive in the next generation. Symbolically, in each generation $h$, we can assign to each cell, $(p, q)$, a state $State(p, q, h)$, which takes the value 1 if the cell is live and 0 if it is dead. At the beginning of the game, when $h$ = 0, the pattern – a boundary condition – is specified for every cell in the starting matrix: $State(p, q, 0)$. The patterns of successive generations are then found from:

$$State(p, q, h+1) = \begin{cases} 1 & \text{if } N(p, q, h) \leq 3 \text{ and } \big(3 - State(p, q, h) \leq N(p, q, h)\big) \\ 0 & \text{otherwise} \end{cases} \qquad (2)$$

where

$$N(p, q, h) = \left\{ \sum_{a=p-1}^{p+1} \sum_{b=q-1}^{q+1} State\,(a, b, h) \right\} - State(p, q, h) \qquad (3)$$

The parameter $N$ is the number of live neighbours surrounding a central cell. In (3), the value of the central cell's state is subtracted from the double summation because that value should not be included in the total, which is supposed to include only neighbouring states.

The complete pattern of the Game-of-Life block universe may be reconstructed from algorithms (2) and (3), given the state of all of the cells in the matrix at the beginning of the game. *Notice that the pattern and the algorithms are isomorphic – they are the same mathematical structure.*

An example of how the Game of Life works, based on a simple 7 × 7 matrix, is given in Fig. 1, which shows a starting position at generation $h$ and the subsequent four generations. In each matrix, the value of $N$ is shown for each of the 49 cells. The pattern in this figure is a "glider" (discovered by the British mathematician, Richard Guy (Roberts 2015)), which progresses diagonally across the matrix, completing the cycle in four steps.

To emphasise the block-universe nature of the structure, these same five matrices are displayed in Fig. 2, stacked in order of successive generations with the starting pattern at the bottom.

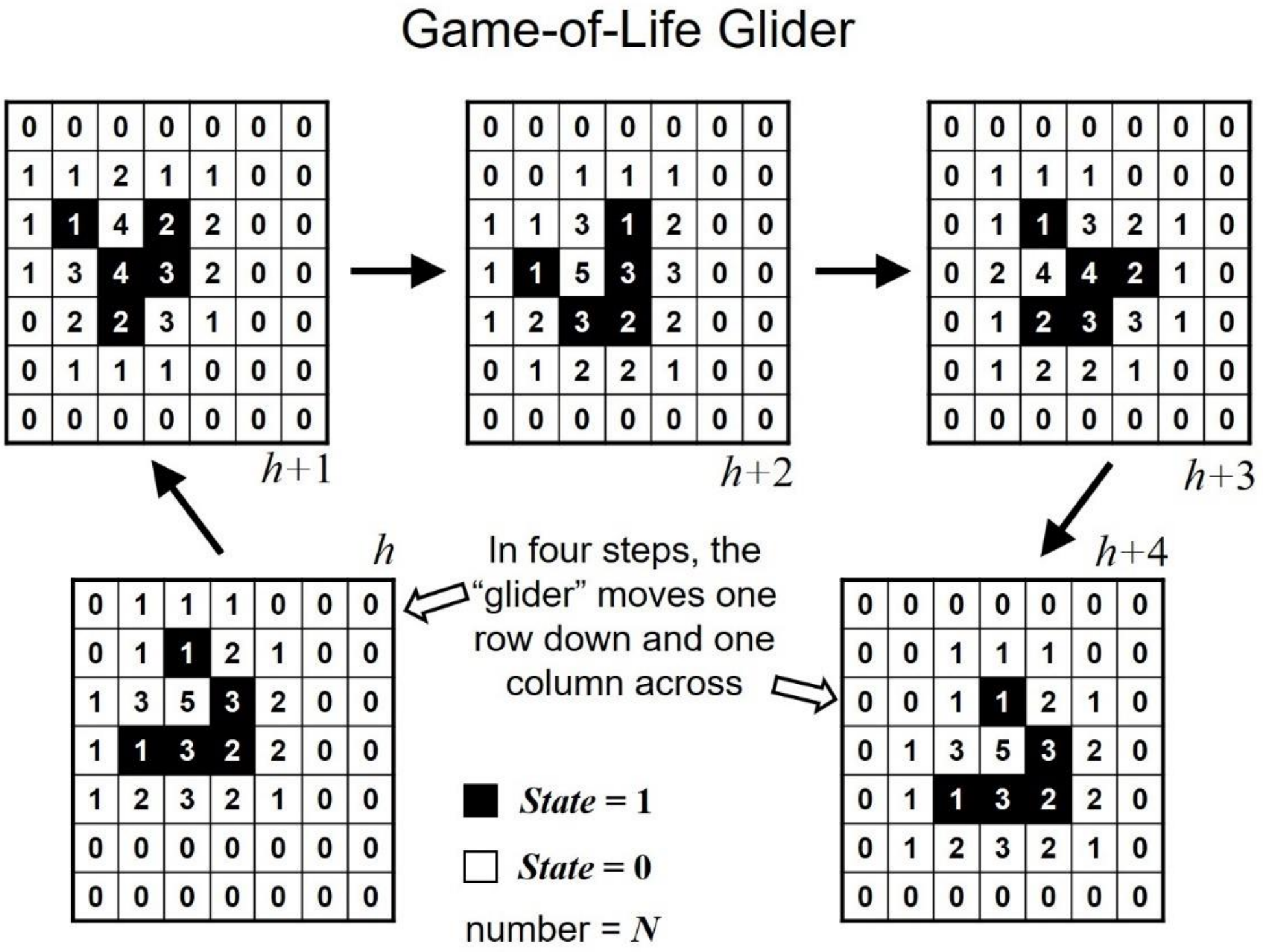

**Fig. 1**
A "glider" in the Game of Life moves downwards and to the right in a cycle of four generations. "Live" cells are shown in black; dead cells are in white. The number in each cell is the number, *N*, of live neighbours surrounding the cell

Remarkably, for structures based upon such a simple algorithm, the Game of Life can support a Universal Turing Machine (UTM) as was first shown by Rendell (2014). The elements of his UTM are shown in Fig. 3. It consists of the rectangular "programmable machine", and a stack that contains both the program and the input data for the program. The stack is diagonal because it relies upon, among other objects, many copies of the diagonally-moving glider depicted in Figs. 1 and 2. With a little licence, one can envisage such a UTM sending probes around its neighbourhood and thereby constructing an internal representation of its matrix environment. When this environment includes itself, then, in a rudimentary sense, we might say that the UTM is self-aware.

If a Game-of-Life player were to pause the execution of a program that included such a "self-aware" UTM and then re-start it at a later time, the UTM would obviously be unaware that the program had been temporarily interrupted. This is to be distinguished from the situation where we build a robot in our laboratory which is aware of its environment, switch it off and then re-start it. Our robot would be aware that it had been temporarily switched off because its environment, such as the laboratory clock, would have changed state while it was unconscious. In the scenario with the Game of Life being paused, however, it is the UTM's *complete universe* that is temporarily halted.

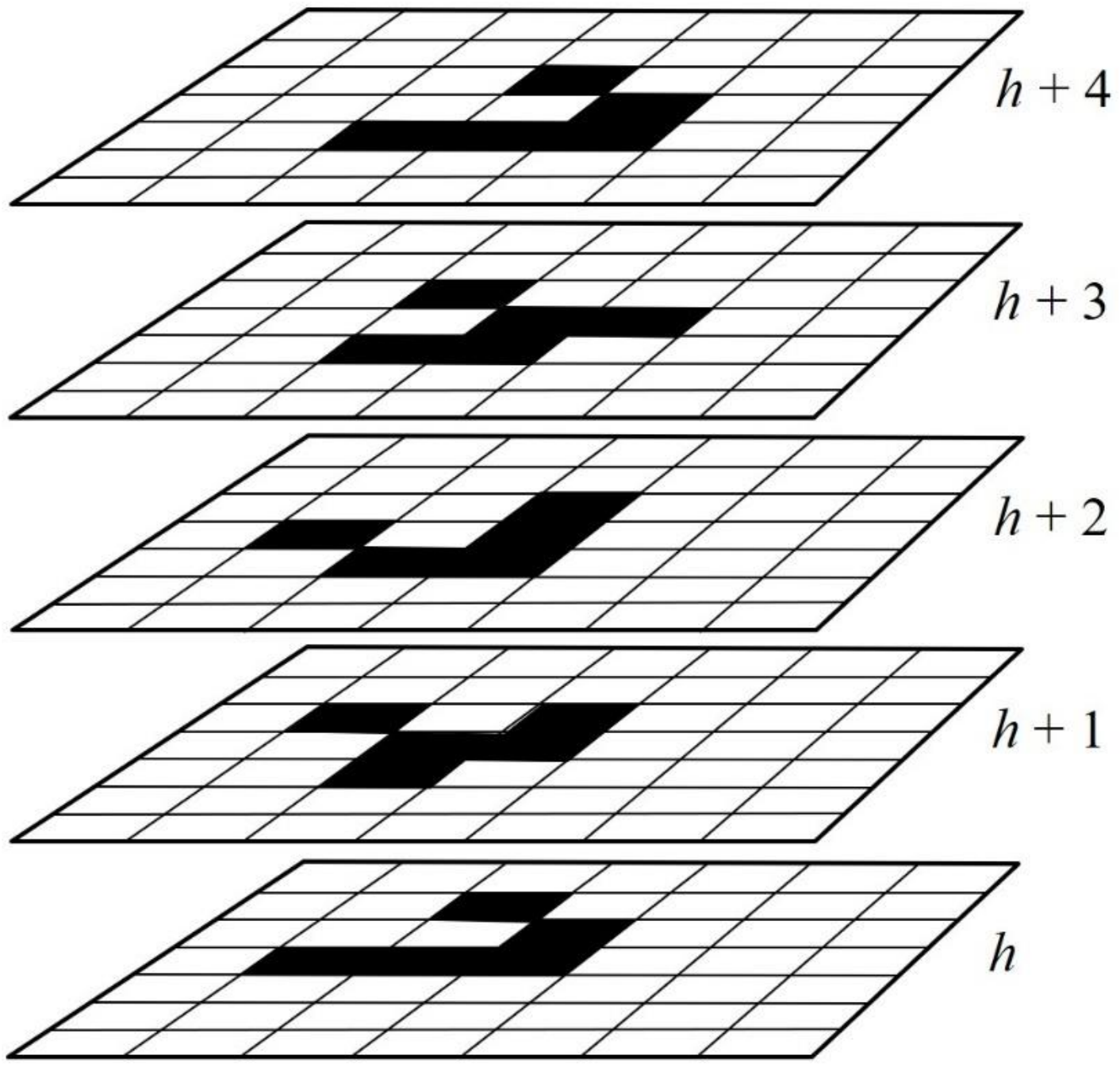

**Fig. 2**
This highlights the block-universe nature of the Game of Life shown in Fig. 1

This simple scenario is intended to illustrate the independence of the pattern – that is, the mathematical structure – of the Game-of-Life block universe from the computer programs that *simulate* it. It is important to appreciate that the UTM in our Game of Life does not spring into self-awareness when the program is started up: the self-awareness of the UTM is a property of the pattern of the particular Game-of-Life block universe in which it finds itself. Its self-awareness is completely independent of the simulation, which has no self-awareness.

Notice, too, that the pattern for the block universe is unique: simulations of it by different computers are different simulations, but they are simulations of a *unique* mathematical structure. In the same way, there may be different simulations of a sphere on different computers, but the *structure* that they are simulating ($x^2 + y^2 + z^2 = constant$) is unique.

## 4. Emergent parameters in a simple mathematical structure

It may be argued that the structure for the Game-of-Life block universe and that for a sphere do not themselves create the three dimensions of the worlds in which their patterns operate. In other words, some may argue that these mathematical structures can only exist by virtue of our own universe, which provides the requisite geometry, and that these mathematical structures are therefore not independent of our universe. However, that perception arises because we used the labels *h*, *x*, *y* and *z*, which are suggestive of our own geometry, for convenience. Fundamentally, there are no labels, only patterns.

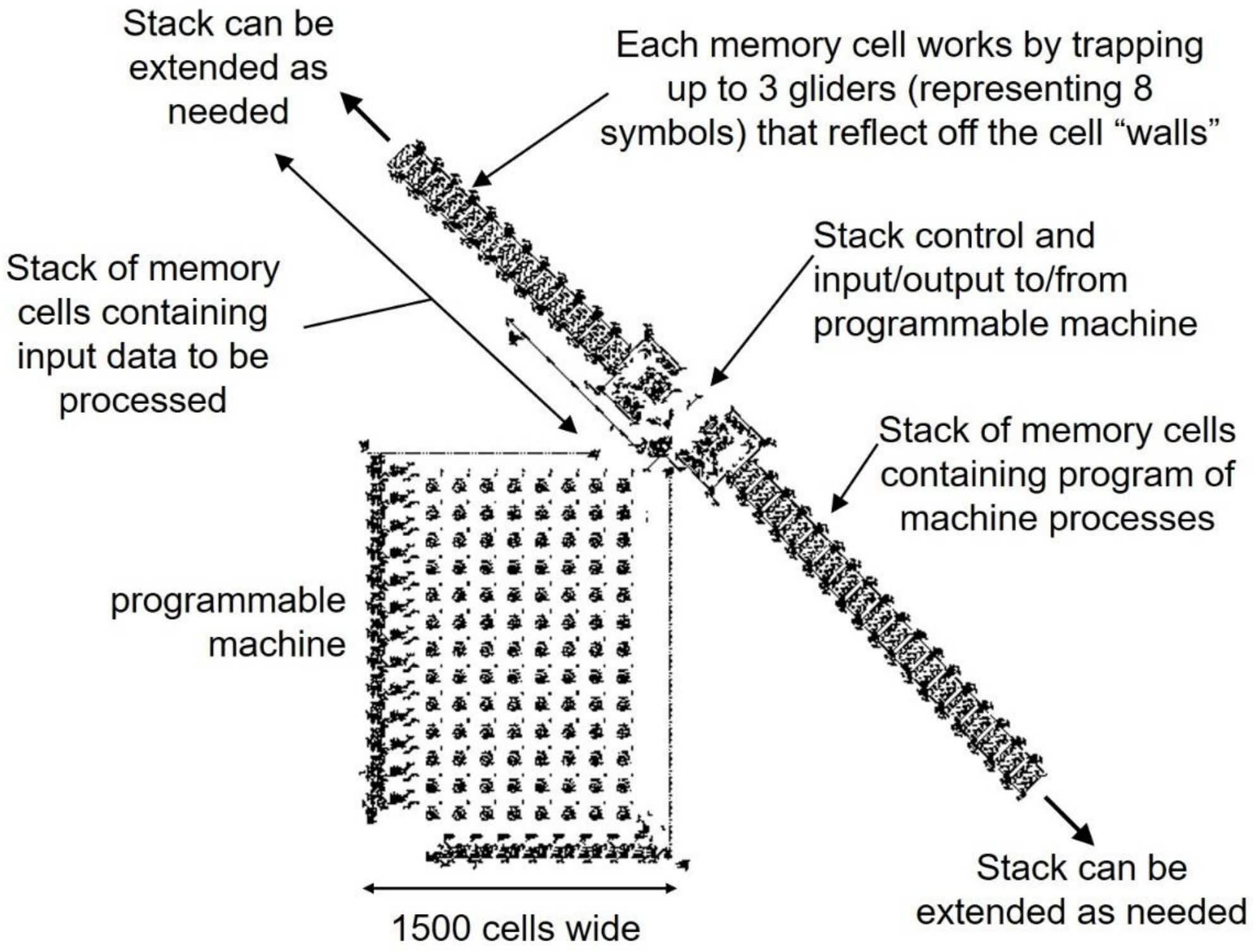

**Fig. 3**
A Universal Turing Machine (UTM) implemented on the Game of Life. Drawing adapted from Rendell (2009)

As a simple illustration of this, take the elementary mathematical structure which is the binary representation of the number 33874822719. This number is expressed in 35 binary digits:

1 1 1 1 1 1 0 0 0 1 1 0 0 0 1 1 0 0 0 1 1 0 0 0 1 1 0 0 0 1 1 1 1 1 1

While there is clearly a pattern within these bits, it is not a particularly "interesting" one. However, since 35 is the product of two prime numbers, the bits can be arranged, in the same order, in a 5 × 7 matrix:

| | | | | | | |
|---|---|---|---|---|---|---|
| 1 | 1 | 1 | 1 | 1 | 1 | 0 |
| 0 | 0 | 1 | 1 | 0 | 0 | 0 |
| 1 | 1 | 0 | 0 | 0 | 1 | 1 |
| 0 | 0 | 0 | 1 | 1 | 0 | 0 |
| 0 | 1 | 1 | 1 | 1 | 1 | 1 |

This arrangement is hardly more interesting than the previous sequence. However, if the digits are arranged in the alternative 7 × 5 matrix, then a more coherent pattern emerges:

| | | | | |
|---|---|---|---|---|
| 1 | 1 | 1 | 1 | 1 |
| 1 | 0 | 0 | 0 | 1 |
| 1 | 0 | 0 | 0 | 1 |
| 1 | 0 | 0 | 0 | 1 |
| 1 | 0 | 0 | 0 | 1 |
| 1 | 0 | 0 | 0 | 1 |
| 1 | 1 | 1 | 1 | 1 |

This pattern lends itself to simple descriptions such as "digits in the 7 × 5 matrix are "1" when they are on the perimeter and "0" otherwise". So, in a Kolmogorov-complexity sense (Li and Vitanyi 2008) it singles itself out as special. Hence, even in this crude example, without using labels, a dimensionality emerges from within the structure, rather than the other way around. We call the parameters that define the dimensionality (in this case, seven rows and five columns) *emergent parameters* – these are defined in Appendix A1.

With the above discussion in mind, our definition of reality may now be refined:

Reality (II)

*Our reality is the complete mathematical structure of our block universe.*

## 5. Fundamental uncertainty implies a parallel-universe model

Let us now look at some of the patterns within our block universe. We may be particularly struck by the outcomes of identical double-slit experiments, each using a single electron. We can arrange for these experiments to be distributed throughout our block universe, including experiments separated by space-like intervals (taking place at the same moment in widely separated laboratories) as well as experiments performed serially in the same laboratory. In each experiment, only a single electron is used, and we note the position of its interaction on the detector screen in every case.

When we look at the outcomes of a very large, but finite, number of experiments, it is clear that, while the position of the interaction on the screen in any single experiment could not have been predicted (it is apparently random), there is nevertheless a pattern that connects the ensemble of outcomes. When the outcomes of each experiment are collected together and plotted on a frequency distribution graph across the detecting screen, the result is a sinc-squared term multiplied by a cosine-squared term: $\left(\frac{\sin\beta}{\beta}\right)^2 \cos^2\alpha$ , where $\alpha$ and $\beta$ are both functions of the distance across the detecting screen. We notice that this pattern matches the absolute square of the value of the probability amplitude along the detecting screen, which we can calculate from the Lagrangians of the quantum electron field summed over time for all of the different trajectories from the slits to the screen.

We notice further that there is always one – and never more than one – interaction between the quantum electron field and the detecting screen. This is surprising because points on the screen are generally space-like separated: after all, if the screen is sufficiently far from the double slits, then the detecting area would have to be kilometres wide. Our surprise stems from the fact that we already noticed that disturbances in quantum fields transmit at a finite speed, so that there is no way for information about an interaction at one point on the screen to be relayed by the quantum field to the other parts of the screen to prevent a duplicate interaction.

Fig. 4 shows two possible positions for the electron to be detected at the screen. The most important point illustrated in this figure is that the quantum electron field (and all of the other quantum fields which have no significant bearing on this particular experiment) is the same at the instant at which the electron is detected, no matter where it is detected. So, it would not have been possible to analyse the quantum electron field right up until the moment of detection and see that there is a configuration that means that the electron will be detected *here* rather than *there*.

There appear to be only two ways in which such experiments, which we have distributed both in space and in time, can be unpredictable individually and yet be connected through a common pattern, such as the sinc-squared–cosine-squared term in this example. The first way is that, in each experiment, the point on the screen where the electron is detected is determined randomly, but with the randomness being weighted according to the absolute square of the probability amplitude derived from summing the Lagrangians of the quantum electron field over time for all possible trajectories between the electron gun and the detecting screen (see Fig. 4). Of course, this is a conventional formulation of quantum mechanics.

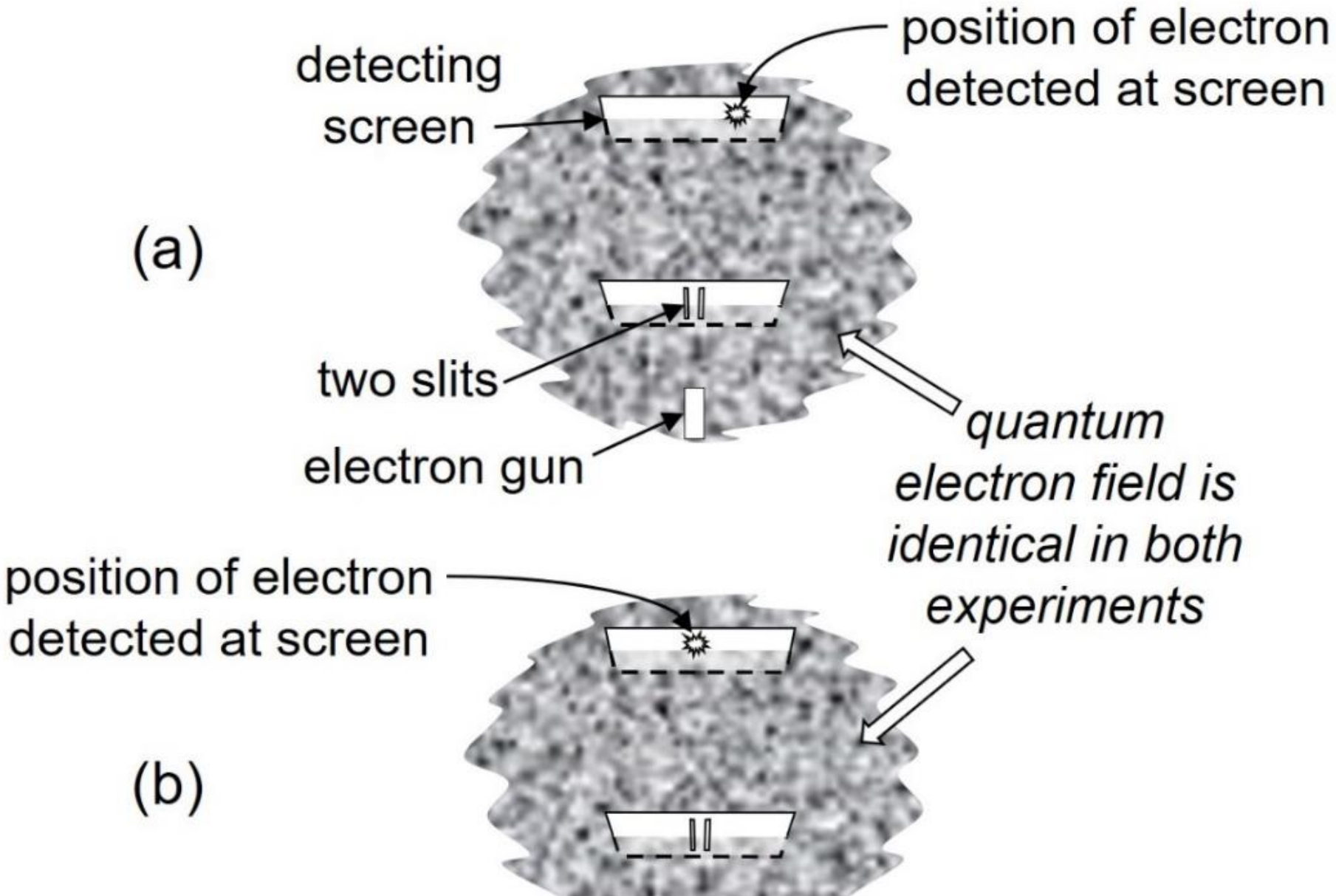

**Fig. 4**
The quantum electron field is identical in both experiments, even though there are two different outcomes (with the electron being detected at different positions on the screen). Therefore, the outcome cannot be predicted from the quantum field

However, since this is a probabilistic explanation, it requires a mechanism that effectively throws the dice for each experiment. The difficulty with this explanation is that no mathematical algorithm or random-number generator can supply the necessary unpredictability, because, by definition, such algorithms are ultimately predictable. Truly random sequences exist in mathematics, of course, such as the digits in a Chaitin halting probability, but such sequences are unknowable in principle – they cannot be computed – and so cannot be candidates for a random-number generator. The only exception is random-number generation using quantum processes (Herrero-Collantes and Carlos Garcia-Escartin 2017), but that would be a logical circularity: you cannot fundamentally explain quantum randomness by saying that it is based upon quantum randomness!

Fundamentally, the problem with the probabilistic explanation comes back to the block universe: if you picture our universe as a block of spacetime embedded throughout with outcomes of quantum interactions, what is it about the underlying pattern of our universe that determines which of the several or many possible outcomes appears at each embedded interaction? Simply to put it all down to probability might at least suggest that the model is incomplete.

The only other way for a pattern to emerge across the individually unpredictable outcomes of identical experiments widely spread in time and space is through a parallel-universe model. A rudimentary example of the particular model used in this paper is described in Appendix A2 as a “Toy Multiverse”. The three key features of the parallel-universe model are (1) there is a finite number of discrete, parallel block universes, in contrast to the branching structure of the Many Worlds Interpretation (MWI); (2) the universes are independent of each other (they do not interact); and (3) their numbers are distributed according to the Born probabilities of the quantum outcomes that they each contain.

Hence, for example, in the case of a Stern-Gerlach experiment with two possible outcomes, one with a 75% chance of occurring and the other with a 25% chance of occurring, there are three times as many universes in the multiverse containing the high-probability result than there are universes that contain the low-probability result. If the same experiment is repeated a million times, this will mean that, in the multiverse, most of the universes which feature the sequence of a million experiments will contain approximately 750,000 of the high-probability results and 250,000 of the low-probability results.

While such a hypothesis may be regarded by some as preposterous in its plethora of universes, others who accept that our universe is ultimately purely a pattern see no fundamental reason why the pattern may not be repeated, albeit on an unimaginable (but always finite) scale.

## 6. Super-reality within a mathematical superstructure

Our definition of reality applies equally to all of the parallel universes. Since each universe is independent of the others, each reality is confined to its own universe. Hence, in Fig. 4, reality in the block universe containing the quantum electron field configuration shown in experiment (a) includes the electron being detected to the right of the screen, whereas reality in the block universe containing the identical quantum electron field configuration at the moment of detection in experiment (b) includes the electron being detected near the centre of the screen.

A definition of reality that encompasses parallel block universes is:

<u>Reality (III)</u>

*Reality within a parallel, block universe is the complete mathematical structure of that parallel, block universe.*

Equally, since each universe is itself a mathematical structure:

<u>Reality (IV)</u>

*Reality within a mathematical structure is the complete definition of that structure.*

To clarify this final statement, an incompletely defined structure would, for example, be one in which the value of a function of a parameter can be, say, either one or zero, with the actual value being left unspecified. The parallel with quantum uncertainty is clear: there is a unique, definite outcome for every quantum event in our block universe, as may be seen when viewing the event in retrospect. The uncertainty arises because such an outcome cannot be determined from the mathematical structure: it is unpredictable.

From the above example using parallel universes to account for the outcomes of the Stern-Gerlach experiment, the complete pattern, or mathematical structure, of any given block universe is apparently not unique. Indeed, since each block universe is full of events with two or more possible outcomes, there will be, in general, many exact copies of any given block universe. This may be seen by inspecting equation 6 and Fig. 13. In that figure, each of the block universes has several exact duplicates, and the numbers of such duplicates will increase with the total number of events and outcomes in the universes.

As a concrete example of this, consider the three block universes of type $i = 2$ shown in Fig. 13, containing outcomes $A_2$, $C_1$, $B_1$. In Fig. 10, the ratios of probabilities of outcomes $C_1 : C_2 : C_3$ are given as 1:3:1. This might lead us to think that there is one solitary universe with outcome $C_1$ and three universes containing outcome $C_2$. However, we see that there are, in fact, three universes of type $i = 2$ in Fig. 13, rather than one solitary universe. If we try to reduce the number of universes in the Toy Multiverse by a factor of three, so that the population of type $i = 2$ is reduced to one solitary universe, then the total number in the $A_2$ branch would also have to be reduced from 30 to 10. However, by the same token, the total number in the $A_1$ branch would need to come down from 10 to $3\frac{1}{3}$, which fails the requirement that the number of universes is necessarily integral. So, in general, there must be a (very large) number of indistinguishable copies of any block universe in the multiverse.

Now, the block universe is a *unique* mathematical structure, just as we observed earlier that a sphere, or, for that matter, the sequence of the first hundred prime numbers, is a unique mathematical structure. Indeed, the adjective “unique” is superfluous in such a context. If we remember that the block universe is purely a pattern – a mathematical structure – then it is just as wrong to regard it as having an indistinguishable copy as it is to think of the sequence of primes from 2 to 97 as having an indistinguishable copy. The structure is the structure!

However, the above paragraph is clearly in conflict with the preceding one which states that there must generally be many indistinguishable copies of any block universe in the multiverse. The difficulty arises because we regard each block universe as an independent mathematical structure. David Deutsch (1997) can claim that the double-slit pattern arises from interference between many universes just after branching/splitting, because quantum interference is a feature of Everett's MWI formulation. However, as stated earlier, the topology of a block universe rules out such branching in our model: the mathematical structures of a multiverse of block universes never overlap, and are therefore independent of each other, and this includes groups of those that are indistinguishable from each other.

The only way for the mathematical structures of the block universes to be unique and yet allow for indistinguishable copies of such structures is for all of the structures to be embedded within a mathematical *superstructure*. Appendix A1 clarifies our terminology of mathematical structures and superstructures and shows how they are related.

In our universe, each quantum field, $\varphi_i$, is a function of location in the block universe, and we may make this dependence explicit by writing it as $\varphi_i(x, y, z, t)$. From equation (1), our universe $U$ is dependent upon the same parameters: $U(x, y, z, t)$. Using $M$ to represent the mathematical superstructure of our multiverse, its pattern may then be written symbolically as

$$M(\theta) \Rightarrow \{U_n(x_n, y_n, z_n, t_n): 1 \leq n \leq N\} \quad (4)$$

where, as is stated in Appendix A1, "⇒" is to be read as "contains the following set of embedded structures". The limit, $N$, is the total (finite) number of parallel universes in our multiverse. The reason for considering $N$ as finite is discussed in the second paragraph of Appendix A2.

As in the Toy Multiverse of Appendix A2, there must be at least one parameter $\theta$ that determines the quantum-mechanical distribution of the $N$ *completely independent* universes, $U_n$, embedded in the superstructure of our multiverse.

Equation (4) is effectively the same as equation (A1) that describes the Toy Multiverse of Appendix A2, although, in our multiverse, the number $N$ is, of course, unimaginably greater than the 40 universes of the Toy Multiverse. The case for assigning a unique set of parameters, $\{x_n, y_n, z_n, t_n\}$, to each universe, $U_n$, is also more subtle than the one we used for the Toy Multiverse. Indeed, the alternative, where all of the universes in our multiverse would share a common set of parameters, $\{x, y, z, t\}$, was actually put forward by Aguirre and Tegmark (2011).

They suggested that the parallel universes (corresponding to our $U_n$) in their Level III Multiverse are similar to, or identical copies of, our own Hubble volume, distributed far across the cosmos. Their model fails because, as McKenzie argues (2017), the eigenstate of any given parallel universe would extend throughout the whole cosmos – that is, including regions of space that are receding from each other at superluminal speeds. Since this applies to all parallel universes, many of which are in mutually orthogonal eigenstates, the scenario of a common set of parameters is ruled out because of this potential clash of eigenstates.

Notice that, since the universes $U_n$ are embedded within the superstructure, $M$, it is permissible to have duplicate – that is, identical – universes. Duplicate universes are identical to each other when viewed from within each one – that is, using only the unique set of emergent parameters belonging to each individual universe. (The emergent parameters may be regarded as the set of parameters, $\{x_n, y_n, z_n, t_n\}$, upon which each universe, $U_n$, depends.) However, at the level of the superstructure, they are distinguishable by their "position" in the superstructure (formally, by the parameter(s) $\theta$ and the index $n$ that applies to every universe $U_n$).

From the Reality (IV) statement, since each universe, $U_n$, has its own, unique set of emergent parameters, $\{x_n, y_n, z_n, t_n\}$, each universe also has its own, unique reality. Because each set of emergent parameters is unique to each universe, the realities within each universe cannot ever "overlap" in any way. However, since the superstructure $M$ contains all of these parallel universes, they are all part of its reality. In order to distinguish between (1) the separate realities of the individual parallel universes, $U_n$, of which ours is one; and (2) the reality of the superstructure, $M$, which is our multiverse, it is convenient to use the term *super-reality* for reality within the superstructure.

It is natural to ask what this super-reality might "look like". Since we are constrained by the emergent parameters of our own universe, then our own universe is the only part of the super-reality that we can explore in any depth. There may be some limited insight to be gained by asking why super-reality appears to be structured along quantum-mechanical lines rather than those of any other paradigm. However, that approach might just turn out to be equivalent to asking why we were born in this century and not 200 years ago.

In the final analysis, no picture of the higher reality can ever be verified by checking it directly. By definition, the vast structure of super-reality that is inaccessible to our own universe exists only for that super-reality, lying forever hidden beyond our own horizon.

## 7. Conclusion

In this paper, we support the case proposed by others that our universe is, fundamentally, a pattern, a mathematical structure. This suggests a relatively simple definition: reality within a mathematical structure is the complete definition of that structure. The idea that our universe is ultimately a pattern may also make it easier to imagine the pattern being extended to include a whole multiverse of parallel universes (we argue that only such a multiverse can account for the absolute randomness of quantum outcomes).

However, such a picture presents us with a dilemma: in order to account for observed probabilities of quantum outcomes, there must be many identical copies of universes with particular outcomes. But this means that there must be many identical copies of particular mathematical structures. The difficulty is that every mathematical structure is unique – there cannot be two identical mathematical structures any more than there can be two identical sets of the first ten prime numbers – there is only one such set. There can be duplicate *representations* of a set or of a mathematical structure, but every structure is unique.

The resolution of the dilemma is that the structure of every parallel universe must be embedded within – must be part of – a larger pattern, which we call a mathematical superstructure. This means that there can be groups of identical universes which can be distinguished from each other (since they occupy different "parts" of the pattern of the superstructure). Thus, each universe is unique from the perspective of the superstructure (and so the requirement that every mathematical structure is unique is not violated) and yet the internal descriptions of each of these universes will be identical, which is one of the requirements of the multiverse explanation of measured quantum probabilities.

In Appendix A1, we note how sets of parameters can emerge from the process of defining a mathematical structure in the simplest possible terms – we call these *emergent parameters*. Each of the parallel universes in the mathematical superstructure is defined in terms of its own unique set of emergent parameters. The uniqueness of the emergent parameters in each universe means that there can be no interference or overlap between universes. However, the superstructure contains all of the unique sets of emergent parameters of these individual universes, and so every universe is accessible to the superstructure. In addition, the superstructure must contain one or more emergent parameters that are not common to any universe. Such parameter(s) allow all of the universes to be distinguished individually, and they also, presumably, account for the universes being distributed numerically in such a way that leads to the expected ratios of quantum outcomes.

Since the superstructure is a mathematical structure, then the same definition of reality applies to it as to the individual parallel universes embedded within it. This reality, however, is different from the realities within individual universes. While the realities of individual universes can never overlap (since the universes each have a different set of emergent parameters), the reality of the superstructure includes the reality of every embedded universe (since the superstructure includes all of the different sets of emergent parameters of the embedded universes). For the reality of the superstructure, we use the term *super-reality*.

In the end, we seem to have made a philosophical conundrum for ourselves. We have deduced that a mathematical multiverse containing parallel universes must exist as a mathematical superstructure. However, since our universe lacks most of the parameters (including the emergent parameters of the other universes) that are intrinsic to this superstructure, then the vast majority of this superstructure cannot be part of our reality. If we regard the concept of reality as synonymous with that of existence, then we have effectively proved that the superstructure does not exist, despite our earlier conclusion that it does!

Of course, it is a false conundrum, and it may be resolved by adopting a wider viewpoint. In Fig. 9, a UTMA is running in the program of each of the five UTMs, namely UTM1 – UTM5. Each UTMA will deduce, from its narrow viewpoint, that the superstructure Game of Life, which is running the five UTMs, does not exist. However, from the viewpoint of the superstructure Game of Life, all five of the Games of Life embedded within its structure most certainly do exist. From within our own universe, the superstructure of the multiverse will remain forever hidden below our horizon, and the super-reality will be no more than a metaphysical curiosity. If we ever wish to glimpse and understand the super-reality that lies beyond our horizon, then we shall have to elevate our perspective accordingly.

## APPENDIX A1: MATHEMATICAL SUPERSTRUCTURES AND SUPER-REALITY

Fig. 5 is the binary representation of a number that, written in decimal, would contain nearly 1000 digits. In binary, the number contains 3293 bits, and so it can be displayed either as an 89 × 37 matrix or a 37 × 89 matrix, with the sequence of bits in either case beginning at the top-left corner and ending at the bottom right. We have chosen the latter arrangement for the figure.

```
10000000000000000000000000000000000000000000000000000000000000000000000000000000000000001
00000000000000000000000000000000000000000000000000000000000000000000000000000000000000000
00000000000000000000000000000000000000000000000000000000000000000000000000000000000000000
00000000000000000000000000000000000000000000000000000000000000000000000000000000000000000
00000000000000000000000000000000000000000000000000000000000000000000000000000000000000000
00000000000000000000000000000000000000000000000000000000000000000000000000000000000000000
00000000000000000000000000000000000000000000000000000000000000000000000000000000000000000
00000000000000000000000000000000000000000000000000000000000000000000000000000000000000000
00000000000000000000000000000000000000000000000000000000000000000000000000000000000000000
00000000000000000000000000000000000000000000000000000000000000000000000000000000000000000
00000000000000000000000000000000000000000000000000000000000000000000000000000000000000000
00000000000000000000000000000000000000000000000000000000000000000000000000000000000000000
00000000000000000000000000000000000000000000000000000000000000000000000000000000000000000
00000000000000000000000000000000000000000000000000000000000000000000000000000000000000000
00000000000000000000000000000000000000000000000000000000000000000000000000000000000000000
00000000000000000000000000011111000000000000000000000000111110000000000000000000000000000
00000000000000000000000000010001000000000000000000000000100010000000000000000000000000000
00000000000000000000000000010001000000000000000000000000100010000000000000000000000000000
00000000000000000000000000010001000000000000000000000000100010000000000000000000000000000
00000000000000000000000000010001000000000000000000000000100010000000000000000000000000000
00000000000000000000000000010001000000000000000000000000100010000000000000000000000000000
00000000000000000000000000011111000000000000000000000000111110000000000000000000000000000
00000000000000000000000000000000000000000000000000000000000000000000000000000000000000000
00000000000000000000000000000000000000000000000000000000000000000000000000000000000000000
00000000000000000000000000000000000000000000000000000000000000000000000000000000000000000
00000000000000000000000000000000000000000000000000000000000000000000000000000000000000000
00000000000000000000000000000000000000000000000000000000000000000000000000000000000000000
00000000000000000000000000000000000000000000000000000000000000000000000000000000000000000
00000000000000000000000000000000000000000000000000000000000000000000000000000000000000000
00000000000000000000000000000000000000000000000000000000000000000000000000000000000000000
00000000000000000000000000000000000000000000000000000000000000000000000000000000000000000
00000000000000000000000000000000000000000000000000000000000000000000000000000000000000000
00000000000000000000000000000000000000000000000000000000000000000000000000000000000000000
00000000000000000000000000000000000000000000000000000000000000000000000000000000000000000
00000000000000000000000000000000000000000000000000000000000000000000000000000000000000000
00000000000000000000000000000000000000000000000000000000000000000000000000000000000000000
10000000000000000000000000000000000000000000000000000000000000000000000000000000000000001
```

**Fig. 5**
This is a 3293-bit binary number with the sequence of bits beginning with the "1" in the top left-hand corner and finishing in the bottom right-hand corner

Let $r$ and $s$ be the number of rows and columns in the matrix, so that $r = 37$ and $s = 89$ in Fig. 5. With this arrangement, the symmetries of the figure are clear, and, in particular, two rectangles, each identified by a perimeter of "1"s, stand out against a background of "0"s. These features would make it easy to reproduce the figure from a simple algorithm based upon $r$ and $s$. In the case of the alternative arrangement, where there are no coherent embedded shapes ($r = 89$ and $s = 37$), the algorithm would be longer.

It is in this way that the parameters $r$ and $s$ may be said to *emerge* – thus, they may be regarded as *emergent parameters*. In practice, because of the limited space on the page, Fig. 5 is so simple that it might be reproduced by an even shorter algorithm not based upon $r$ and $s$. However, for significantly large matrices in which are embedded a greater variety of shapes, the emergent parameters become key to the shortest reproducing algorithms.

Shorter algorithms are associated with lower Kolmogorov complexity (Li and Vitanyi 2008), and it is in this sense that the parameters *r* and *s* emerge naturally.

A characteristic of the type of simple mathematical object displayed in Fig. 5 is that the space in which smaller objects are embedded (such as the two rectangles) is mutually shared. For instance, if the centres of the two rectangles in Fig. 5 had been separated in the *s* direction by only one bit rather than 29 bits, the two rectangles would share some common space.

```
10000000000000000000000000000000000000000000000000000000000000000000000000000000000000001
00000000000000000000000000000100000000000000000000000000000100000000000000000000000000000
00000000000000000000000000000100000000000000000000000000000100000000000000000000000000000
00000000000000000000000000000100000000000000000000000000000100000000000000000000000000000
00000000000000000000000000000100000000000000000000000000000100000000000000000000000000000
00000000000000000000000000000100000000000000000000000000000100000000000000000000000000000
00000000000000000000000000000100000000000000000000000000000100000000000000000000000000000
00000000000000000000000000000000000000000000000000000000000000000000000000000000000000000
00000000000000000000000000000000000000000000000000000000000000000000000000000000000000000
00000000000000000000000000000000000000000000000000000000000000000000000000000000000000000
00000000000000000000000000000100000000000000000000000000000100000000000000000000000000000
00000000000000000000000000000100000000000000000000000000000100000000000000000000000000000
00000000000000000000000000000000000000000000000000000000000000000000000000000000000000000
00000000000000000000000000000000000000000000000000000000000000000000000000000000000000000
00000000000000000000000000000000000000000000000000000000000000000000000000000000000000000
00000000000000000000000000000100000000000000000000000000000100000000000000000000000000000
00000000000000000000000000000100000000000000000000000000000100000000000000000000000000000
00000000000000000000000000000000000000000000000000000000000000000000000000000000000000000
00000000000000000000000000000000000000000000000000000000000000000000000000000000000000000
00000000000000000000000000000000000000000000000000000000000000000000000000000000000000000
00000000000000000000000000000100000000000000000000000000000100000000000000000000000000000
00000000000000000000000000000100000000000000000000000000000100000000000000000000000000000
00000000000000000000000000000000000000000000000000000000000000000000000000000000000000000
00000000000000000000000000000000000000000000000000000000000000000000000000000000000000000
00000000000000000000000000000000000000000000000000000000000000000000000000000000000000000
00000000000000000000000000000100000000000000000000000000000100000000000000000000000000000
00000000000000000000000000000100000000000000000000000000000100000000000000000000000000000
00000000000000000000000000000000000000000000000000000000000000000000000000000000000000000
00000000000000000000000000000000000000000000000000000000000000000000000000000000000000000
00000000000000000000000000000000000000000000000000000000000000000000000000000000000000000
00000000000000000000000000000100000000000000000000000000000100000000000000000000000000000
00000000000000000000000000000100000000000000000000000000000100000000000000000000000000000
00000000000000000000000000000100000000000000000000000000000100000000000000000000000000000
00000000000000000000000000000100000000000000000000000000000100000000000000000000000000000
00000000000000000000000000000100000000000000000000000000000100000000000000000000000000000
00000000000000000000000000000100000000000000000000000000000100000000000000000000000000000
10000000000000000000000000000000000000000000000000000000000000000000000000000000000000001
```

**Fig. 6**
This is a 3293-bit binary number, similar, but not identical, to that in Fig. 5

In contrast, now consider the matrix in Fig. 6. This figure again features a 37 × 89 matrix representation of a number expressed in 3293 bits, and, again, it contains two shapes, this time in the form of two vertical lines of "1"s, each punctuated by five strings of three "0"s. If the matrix had been arranged in the alternative 89 × 37 arrangement, then no such simple embedded shapes would emerge.

The bits in each vertical line, taken sequentially, are:

1 1 1 1 1 1 0 0 0 1 1 0 0 0 1 1 0 0 0 1 1 0 0 0 1 1 0 0 0 1 1 1 1 1 1

which is the 35-bit binary representation of the same number, 33874822719, that we encountered earlier. As before, prime-factor parameters $p$ and $q$ emerge naturally in this sequence, with $p = 7$ and $q = 5$. Notice that, in order to differentiate between the emergent parameters of the large matrix, $r$ and $s$, we have used $p$ and $q$. To distinguish between the two separate lines, we may use subscripts for the emergent parameters, so that $p_1 = 7$ and $q_1 = 5$ refers, say, to the left-hand line and $p_2 = 7$ and $q_2 = 5$ to the right-hand one.

It is necessary to distinguish between the emergent parameters, $r$, $s$, of the 3293-bit number and those of the two 35-bit numbers, $p_1$, $q_1$ and $p_2$, $q_2$, because, unlike the rectangles in Fig. 5, the rectangles that are implicit in the embedded lines of Fig. 6 do not share the same space as the 37 × 89 matrix. The two lines could be drawn adjacent to each other, but the two implicit rectangles would not overlap, because the spaces defined by the three sets of emergent parameters, $\{r, s,\}$, $\{p_1, q_1\}$ and $\{p_2, q_2\}$, are all different.

Symbolically, we may write:

$$Matrix(r, s) \Rightarrow \{Rectangle_1(p_1, q_1), Rectangle_2(p_2, q_2)\}$$

where "⇒" is to be read as "contains the following set of embedded structures".

A more revealing example is shown in Fig. 7, which is a sketch of the starting matrix, generation $h = 0$, of a Game of Life in which there is not one, but five, UTMs of the type shown in Fig. 3.

As before, the Game of Life may be viewed as a stack of matrices, each evolving from the previous one according to the Game-of-Life algorithms (2) and (3), and represented by $State(p, q, h)$. The block universe formed from the complete set of these matrices is shown in Fig. 8, where we have assumed a finite number of generations, terminating in $h_{max}$.

Since a Game of Life is evidently computable as a simulation, it can be computed on any of the UTMs in Fig. 8. Therefore, let all four outside UTMs, that is, UTM1, UTM2, UTM3 and UTM4, be programmed to simulate identical Games of Life. The resulting block Games of Life may be labelled, respectively, $GoL_1(x_1, y_1, t_1)$, $GoL_2(x_2, y_2, t_2)$, $GoL_3(x_3, y_3, t_3)$ and $GoL_4(x_4, x_4, t_4)$. Notice that we have used emergent parameters $x$, $y$ and $t$ in order to distinguish them from those of the Game of Life running the five UTMs, $p$, $q$ and $h$.

Thus, each of the four outside UTMs will contain the same program data (that is, the program for the Game of Life) and the same input data, which will represent the pattern (that is, the mathematical structure) in the zeroth-generation matrix of the Game of Life.

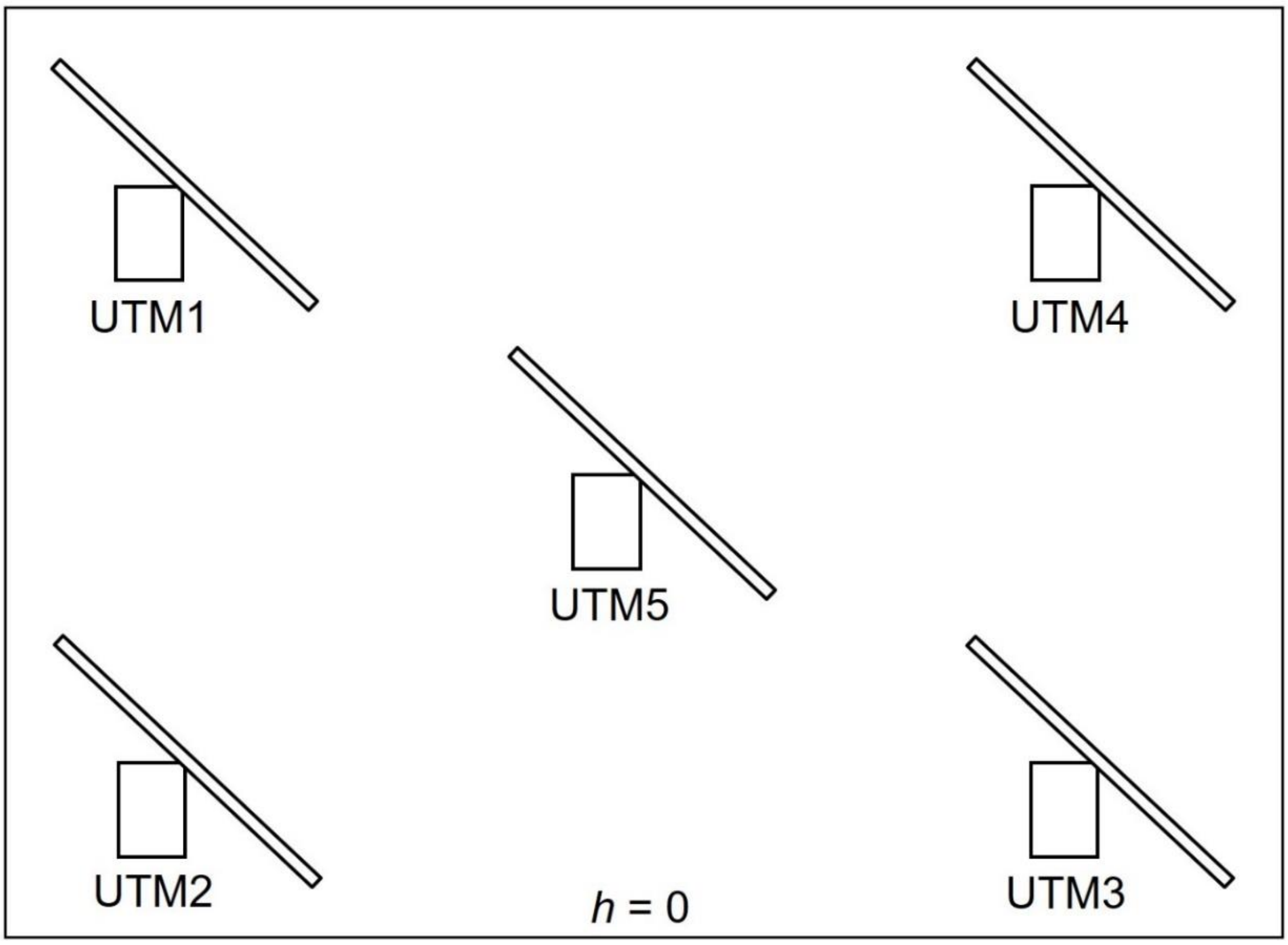

**Fig. 7**
Everything within the outer rectangle is a schematic of the starting matrix of a Game of Life featuring five UTMs, each one of which is, itself, running a Game of Life, $GoL_1$ - $GoL_5$

Now imagine that each of the Games of Life, $GoL_1$ - $GoL_5$, contains within it the instructions for a further UTM of the kind discussed earlier, programmed to explore its environment – let us call it a UTMA. Suppose also that the input data for each of the four outside UTMs contain, in addition to the UTMA, a solitary glider, moving diagonally downwards to the right. Suppose, further, that the central UTM – UTM5 – is again programmed with the Game of Life featuring a UTMA, but that, this time, it also features a solitary glider moving diagonally downwards to the left rather than to the right. The block Game of Life for UTM5 may be labelled $GoL_5(x_5, y_5, t_5)$.

The configuration is summarised in Fig. 9.

Of course, the pattern of the Game of Life which each of the UTMs is running, including the UTMA and glider, cannot be seen simply by looking at the cuboid illustrated in Fig. 8, just as the pattern of $Rectangle(p_1, q_1)$ and $Rectangle(p_2, q_2)$ cannot be seen by inspecting the 37 × 89 matrix in Fig. 6. The patterns of the Games of Life running in the five UTMs are only revealed when they are expressed in their emergent parameters $\{x_n, y_n, t_n\}$.

This is an instance of a general property that, if a mathematical structure, $S$, contains an embedded structure which can be expressed more simply in its own emergent parameters, then $S$ may be regarded as a mathematical *super*structure. Since the Games of Life, $GoL_1$ - $GoL_5$, have emergent parameters that are different from those of the Game of Life running them, then the latter is a superstructure.

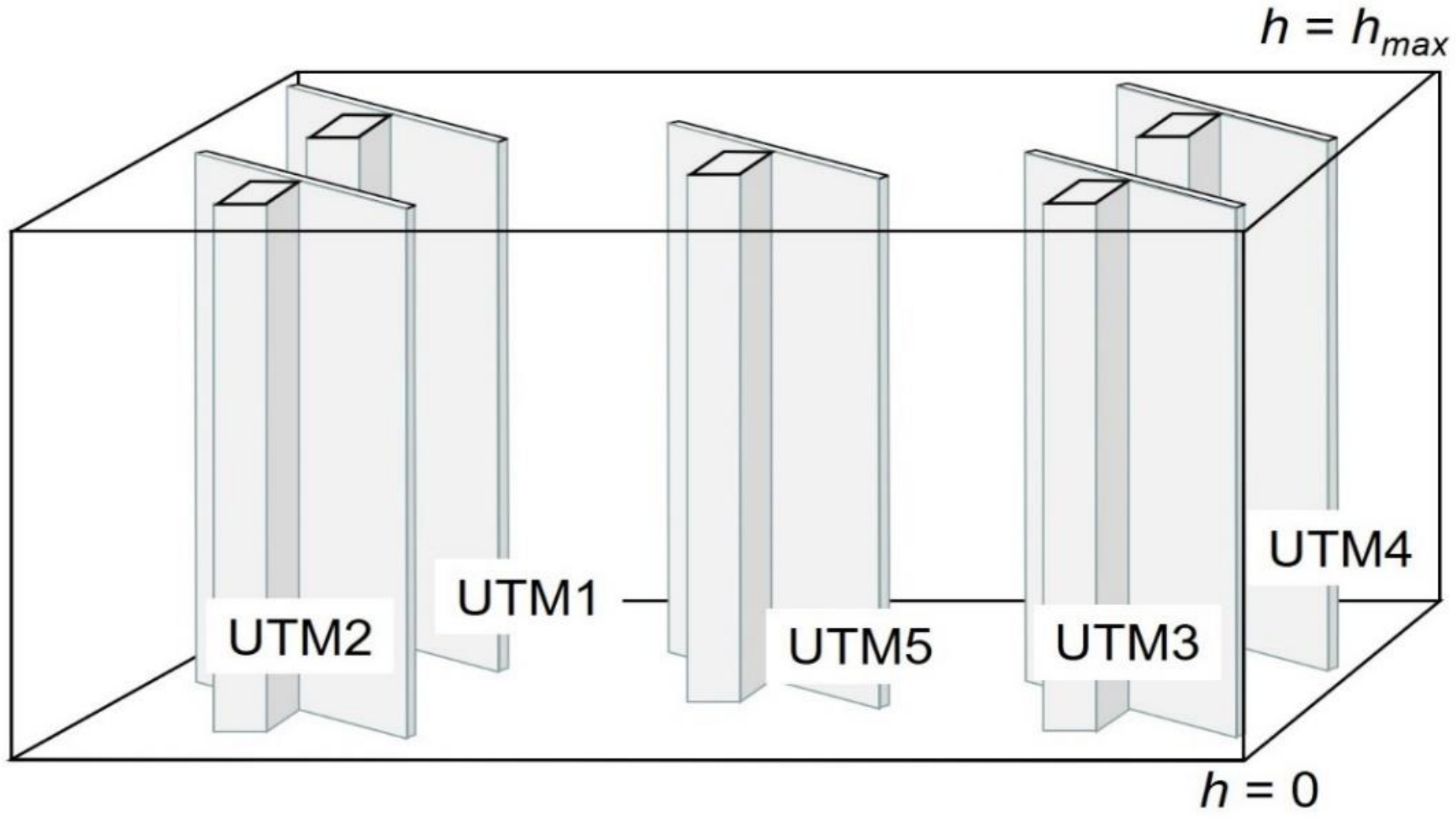

**Fig. 8**
The cuboid shows a complete Game of Life, played out from its starting configuration at $h$ = 0 to the final one at $h$ = $h_{max}$. Five UTMs are running in this Game of Life, each programmed with its own Game of Life, $GoL_1$ - $GoL_5$

The embedded Games of Life running in the five UTMs are independent of each other – the gliders in the Games of Life running in UTM4 and UTM5 can never collide, with each game being played out independently within the respective emergent parameters $\{x_4, y_4, t_4\}$ and $\{x_5, y_5, t_5\}$.

So, in the superstructure Game of Life in Fig. 8, there are five independent realities, one for each of the five Games of Life being played out in the five UTMs. In four of these realities, a glider moves diagonally downwards to the right, and in one of the realities, the glider moves diagonally downwards to the left.

While the five realities are independent of each other, they are part of the reality of the mathematical superstructure in Fig. 8, which we can call a *super-reality*. The complete mathematical structure of each of the five realities is accessible to that super-reality. Symbolically, we may write:

$$State(p, q, h) \Rightarrow \{GoL_n(x_n, y_n, t_n): 1 \leq n \leq 5\}$$

It is in this sense that the Toy Multiverse of Appendix A2, in which 40 universes are embedded, is considered to be a mathematical superstructure. The super-reality of this mathematical superstructure contains, among other things, the 40 mathematical structures that are parallel universes. The realities intrinsic to these 40 universes are completely independent of each other, although they are accessible to the overarching super-reality. As well as the 40 parallel universes, we can conclude that the mathematical superstructure must contain information about the relative probabilities of different quantum outcomes of each of the three quantum events A, B and C, in the Toy Multiverse. This is because the relative numbers of these parallel universes are determined within the mathematical superstructure, and it is these relative numbers that determine the quantum probabilities.

Symbolically, we may write:

$$Multiverse(\theta) \Rightarrow \{U_n(x_n, y_n, z_n, t_n): 1 \leq n \leq 40\} \qquad 5$$

where $\theta$ represents the emergent parameter(s) of the multiverse in which the parallel universes are embedded. We know that the multiverse must contain at least one emergent parameter, because the 40 embedded universes make a pattern – a structure – that follows the rules of quantum mechanics: the frequency distribution of different types of universes is given by equation (A2). The parameter $\theta$ determines the pattern of these 40 universes.

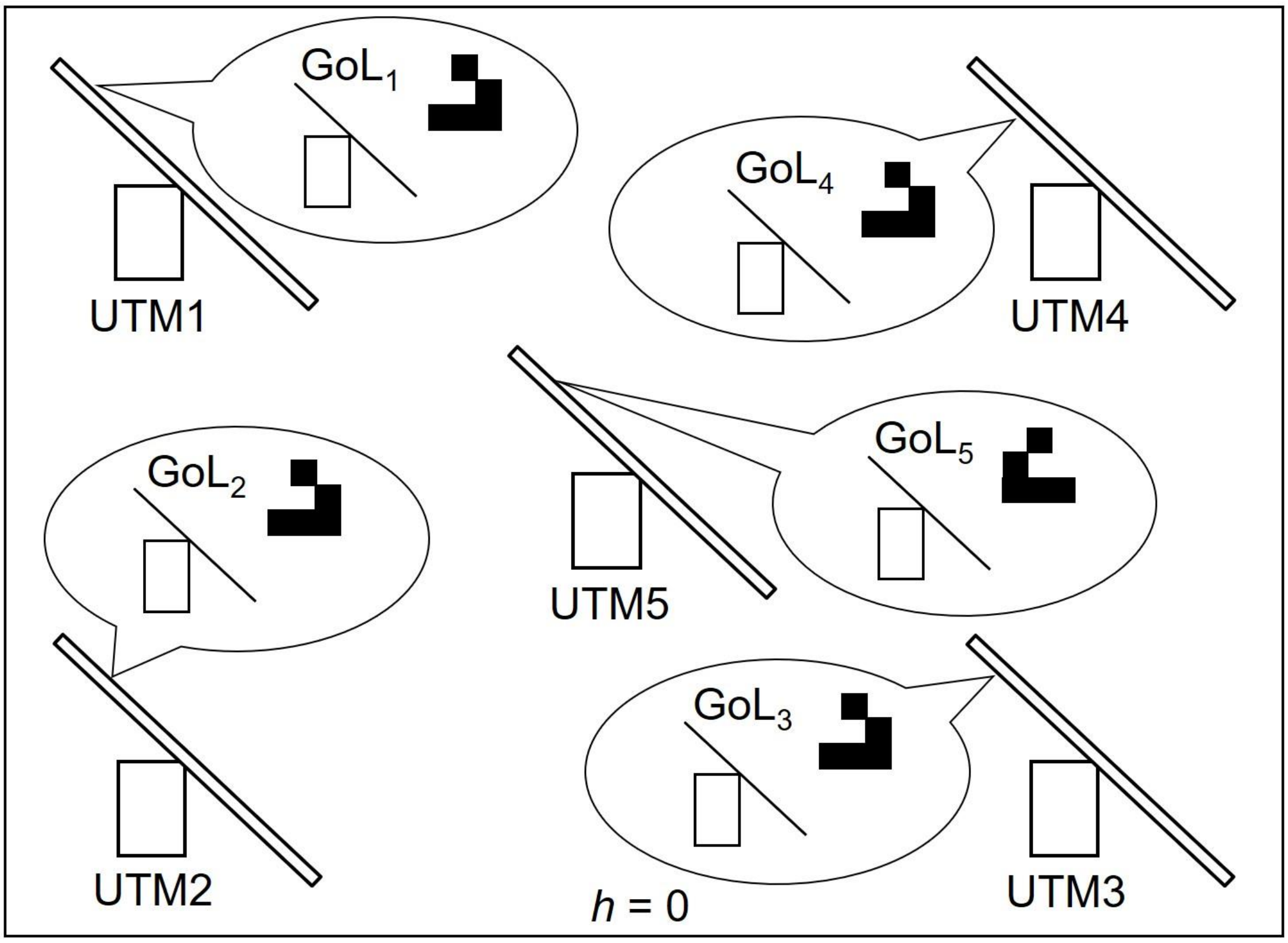

**Fig. 9**
This schematic diagram shows five UTMs, UTM1 – UTM5, running within the Game of Life. Each of these structures contains, in turn, an embedded Game of Life that features a further UTM (called UTMA) and a glider

Note that, in the above equation, each of the 40 universes, $U_n$, depends upon a set of emergent parameters, $\{x_n, y_n, z_n, t_n\}$ which is unique to that universe. The fact that each universe must depend upon a set of parameters which is unique to that universe may be seen by considering the alternative, where all 40 universes share a common set of parameters $\{x, y, z, t\}$. In that case, the 40 universes would merge to form a single structure – effectively, a single, large universe. So, instead of this enlarged universe containing a maximum of three quantum events, which is one of the specifications for the model, it would contain 110 quantum events.

## APPENDIX A2: A TOY MULTIVERSE

While Everett's work was (somewhat sensationally) reinterpreted by DeWitt as invoking parallel universes (DeWitt 1970, DeWitt and Graham 1973), Everett's main aim was to preserve the unitary evolution of the Schrödinger equation without having to incorporate wave function collapse. Strictly, the "parallel universes" of the Many Worlds Interpretation are, topologically, a single structure that branches at every quantum interaction ("measurement"), and it is this topology that is in conflict with that of a block universe, which has no branches. The resolution of this conflict is to replace the branches with separate, discrete, parallel filaments (*i.e.*, universes) with the numbers of filaments/universes in the branches being proportional to the "thicknesses" of the branches (*i.e.*, the Born probabilities of the quantum outcomes that give rise to the individual branches) (see, for instance, Figs. 13 and 14). In such a multiverse of separate, parallel universes, the Schrödinger equation evolves in exactly the same unitary fashion without recourse to wave function collapse, just as Everett intended in his original work.

The features of this model are discussed by McKenzie (2016(a)). The most important of these is that the total number of parallel universes in the multiverse is large but finite. This is shown by McKenzie (2016(c)); essentially, it is because the probability of a given outcome of a quantum event is determined by the ratio of the number of universes containing that particular outcome to the total number of universes containing the quantum event. If the numbers of universes in the ratio were infinite, that ratio would be inconsistent and undefined, whereas the measured probabilities of quantum outcomes are consistent and well defined. At the same time, assuming that all of the branches stem from a common source (so that the parallel universes look increasingly similar to each other towards their origins), then $n_i$, which is the number of universes of type $i$, defined by a given set of quantum outcomes, $k_i$, that result from the quantum events, $j_i$, is given by

$$n_i = N \prod_{j_i} |\langle k_i | j_i \rangle|^2 \qquad 6$$

where $\langle k_i | j_i \rangle$ is the probability amplitude for quantum event $j_i$ to have outcome $k_i$ in a universe of type $i$. $N$ is the total number of parallel universes in the multiverse, so that

$$N = \sum_i n_i$$

It will be seen from these two equations that $N$ is a free parameter, essentially because

$$\sum_{i} \prod_{j_i} |\langle k_i | j_i \rangle|^2 = 1$$

(A simple example to make this equation plausible is given by McKenzie (2016(c).)

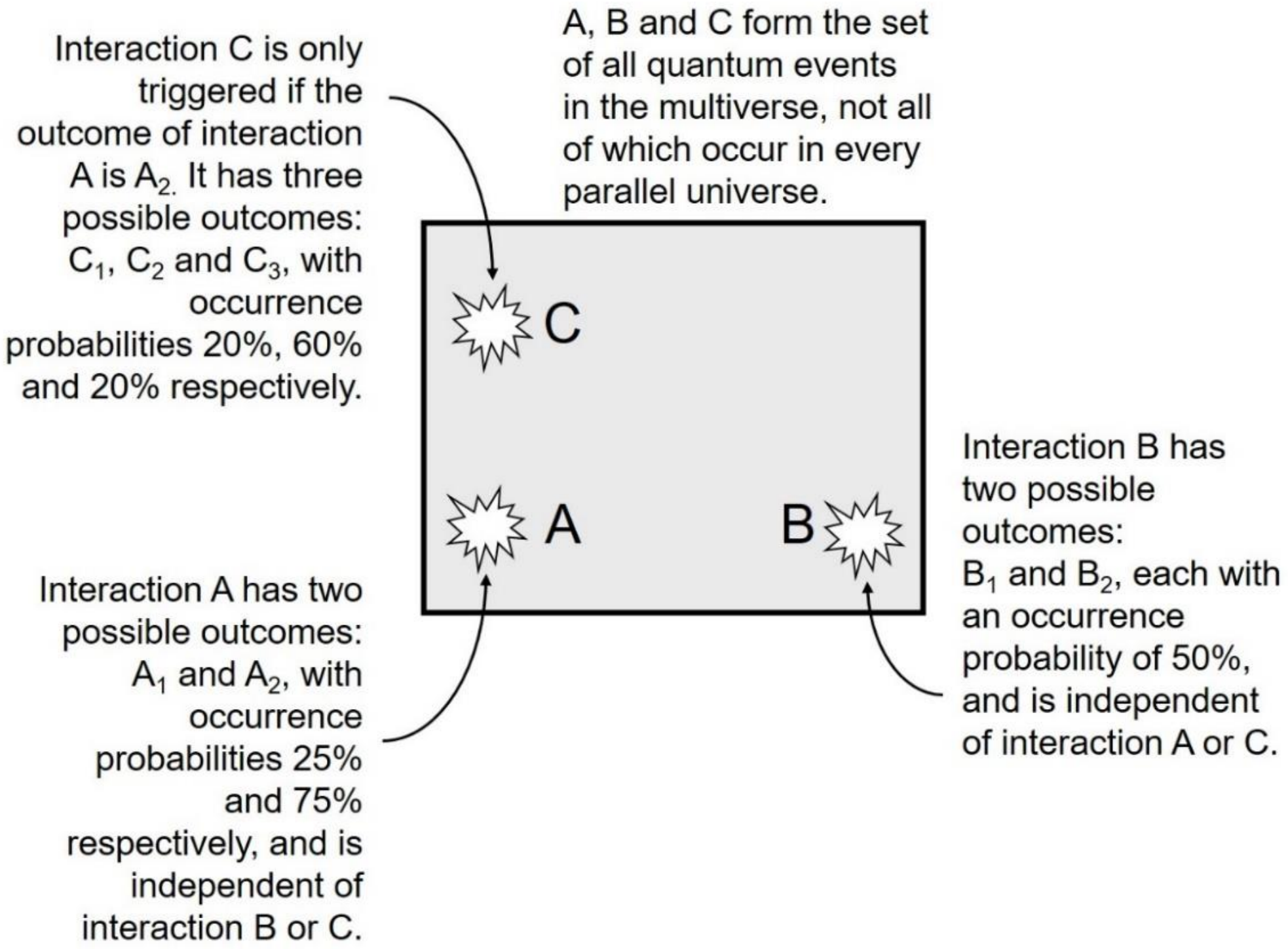

**Fig. 10**
This shows the elements of a simple Toy Multiverse. There is a maximum of only three quantum events, A, B and C in any of the constituent universes

We shall return to the question of the value of $N$, but an illustration may be helpful at this point. Consider the case of a simple Toy Multiverse containing a maximum of three quantum events, A, B and C, as shown in Fig. 10. Event A has two possible outcomes, $A_1$ and $A_2$, with occurrence probabilities of 25% and 75% respectively. If, and only if, the outcome of event A is $A_2$, then an event C is triggered. This has three possible outcomes, $C_1$, $C_2$ and $C_3$, with occurrence probabilities of 20%, 60% and 20% respectively. (The outcomes of C may be thought of as representing the array of possible outcomes across the detector screen in a two-slit experiment.) Event B occurs independently of A and C, and has two possible outcomes, $B_1$ and $B_2$, each with a 50% probability of occurring.

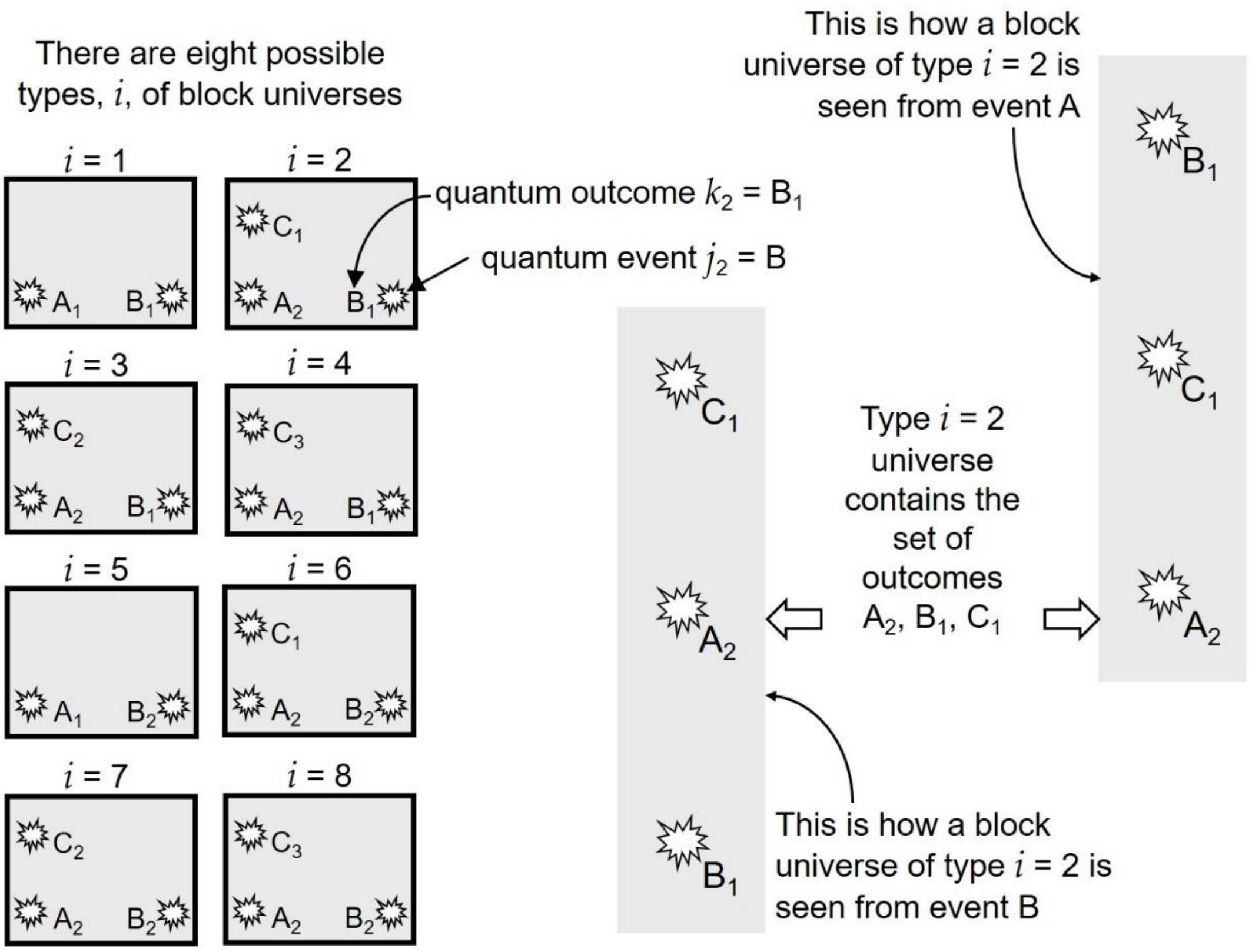

**Fig. 11**
The eight possible outcomes of quantum events A, B and C are shown in the eight boxes. Each box represents a "type" of universe. The outcomes will look different from different perspectives as shown in the two vertical strips to the right of the figure

Fig. 11 shows the eight different possible types, *i*, of universe that can arise in this multiverse. A "type" of universe is defined by the particular set of outcomes arising from the quantum events in that universe. Since the three quantum events in the Toy Multiverse, A, B and C, are not entirely independent (event C is triggered by outcome $A_2$), the total number of different types of universe is not simply the product of the numbers of outcomes of the three different events, $2 \times 2 \times 3$, but, instead, is eight, as shown in Fig. 11. For illustration, the symbols $j_i$ and $k_i$ used in the above equations to represent quantum events and the outcomes of these events, are shown pointing to event B and outcome $B_2$ respectively in the universe of type 2. Notice that the numbering sequence, *i*, used to identify the different types of universe, is arbitrary.

We suppose that event B is sufficiently far from event A that the future light cone of event B does not reach A's world line until after event C, which is on A's world line. So, the sequence of outcomes in a universe of type 2, as seen from event A, is $A_2$, $C_1$, $B_1$. This is illustrated at the far right of Fig. 11 as a vertical strip with the three events drawn in chronological order, starting from the first event at the bottom of the strip. From the vantage point of event B, however, the sequence is $B_1$, $A_2$, $C_1$, also shown in a vertical strip in Fig. 11. Notice that both versions of the vertical strips are represented by the block universe of type $i$ = 2.

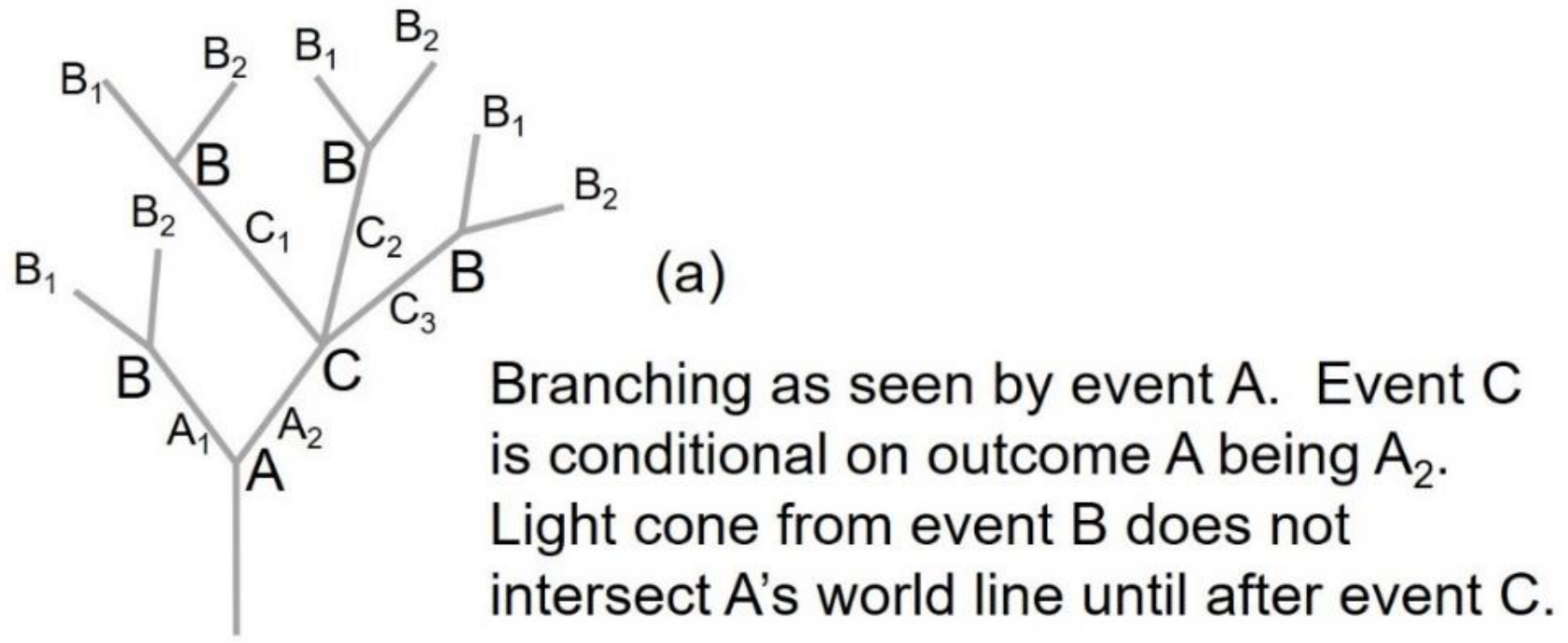

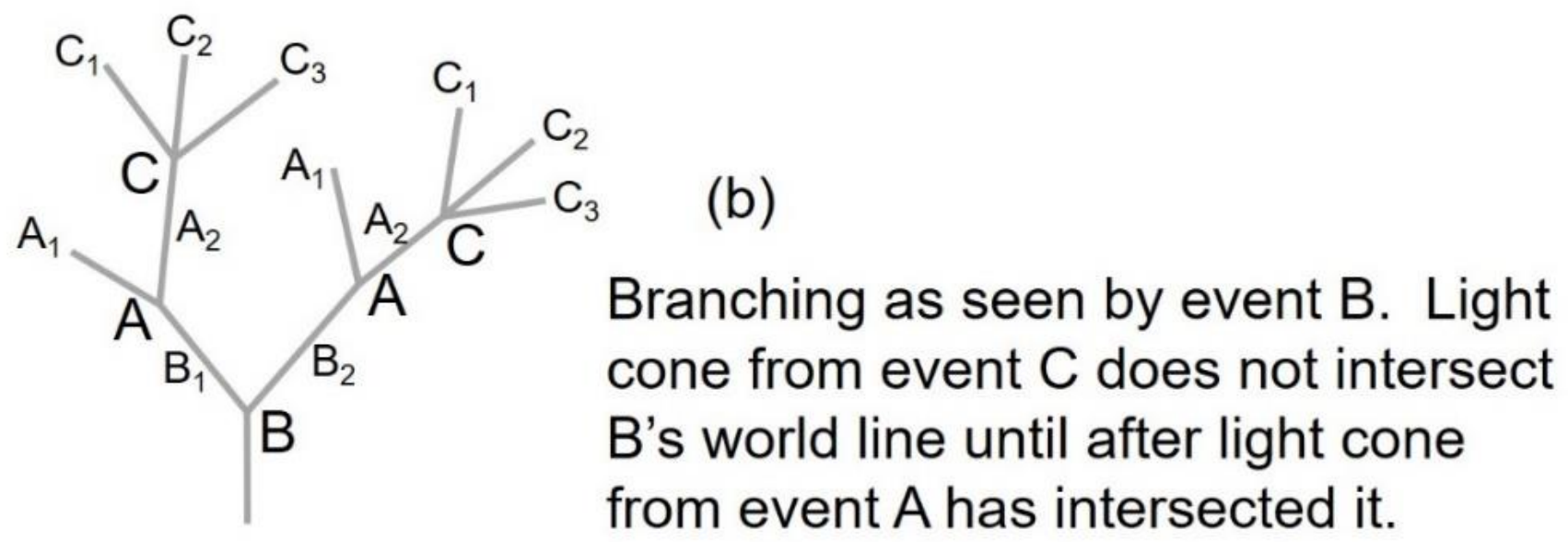

**Fig. 12**
Diagram of the same Toy Multiverse of eight different quantum outcomes as seen in two representations of the MWI. The differences arise because of the different perspectives of the quantum events A and B

Fig. 12 shows how the Toy Multiverse of eight different types of universe may be viewed from event A and also from event B. Each tree is effectively a multiverse according to the Many Worlds Interpretation. The topologies of the two trees are clearly different, although they represent exactly the same events, because the branches have to be drawn from a particular viewpoint, and the number of possible viewpoints increases with the number of events. In this respect, the eight types of block universe in Fig. 11 may be regarded as more fundamental than the MWI picture, because their structure is independent of any viewpoint. This can be more clearly seen by comparing Figs. 13 and 14, which show the same eight types of block universe in both MWI structures.

Fig. 13 corresponds to the MWI tree in Fig. 12(a), where, as described above, the branches have been replaced with separate, discrete filaments with each filament representing one of the eight types of block universe. Fig. 14 corresponds to the MWI tree in Fig. 12(b), and it will be seen that, although the MWI topology of the tree is different from that in Fig. 13, it comprises the very same eight block universes.

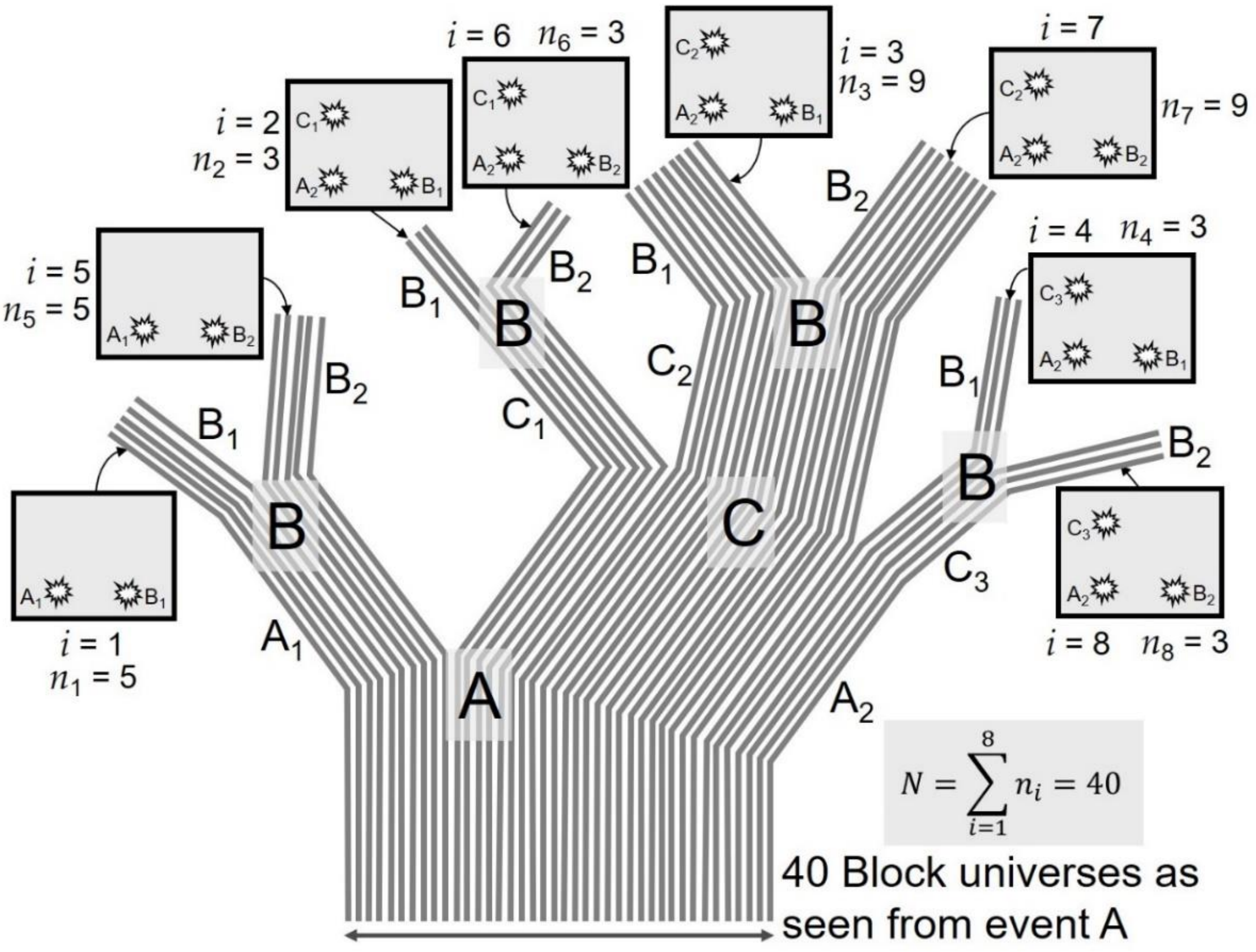

**Fig. 13**
This shows the Toy Multiverse with its eight different types of block universe. Each universe is represented by a long, grey filament. The tree structure corresponds to the MWI structure as viewed from the perspective of event A

A total number of 40 parallel block universes has been used in each of the two trees, which is just enough to demonstrate the probability rules. For instance, in Fig. 13, the trunk of the tree splits into two branches at event A, with 10 filaments/universes in the left branch (outcome $A_1$) and 30 filaments/universes in the right branch (outcome $A_2$). In other words, three times as many universes contain outcome $A_2$ as those that contain outcome $A_1$, which is consistent with the original premise that the relative probabilities of outcomes $A_1$:$A_2$ are 25%:75%.

In Fig. 14, the event A appears twice in the tree, one in each of the two outcome branches of event B. Each of these branches contains 20 universes and it will be seen that, as in Fig. 13, event A splits into a 25%:75% outcome ratio, with 5 universes in the $A_1$ branch and 15 in the $A_2$ branch.

The total number of universes, $N$ = 40, in these two illustrations is the smallest we could have used while maintaining the numbers of universes in all of the branches needed to produce the specified quantum outcome probabilities. Clearly, the total number of universes in the Toy Multiverse can be scaled upwards to any multiple of 40, but that raises the question of what to do if the ratio of two quantum outcomes of an event were irrational: for instance, what would be the number of universes containing, respectively, the outcomes $A_1$ and $A_2$ if the ratio $A_1$:$A_2$, as determined by the Schrödinger equation, were $1/\sqrt{10}$ instead of ⅓ ?

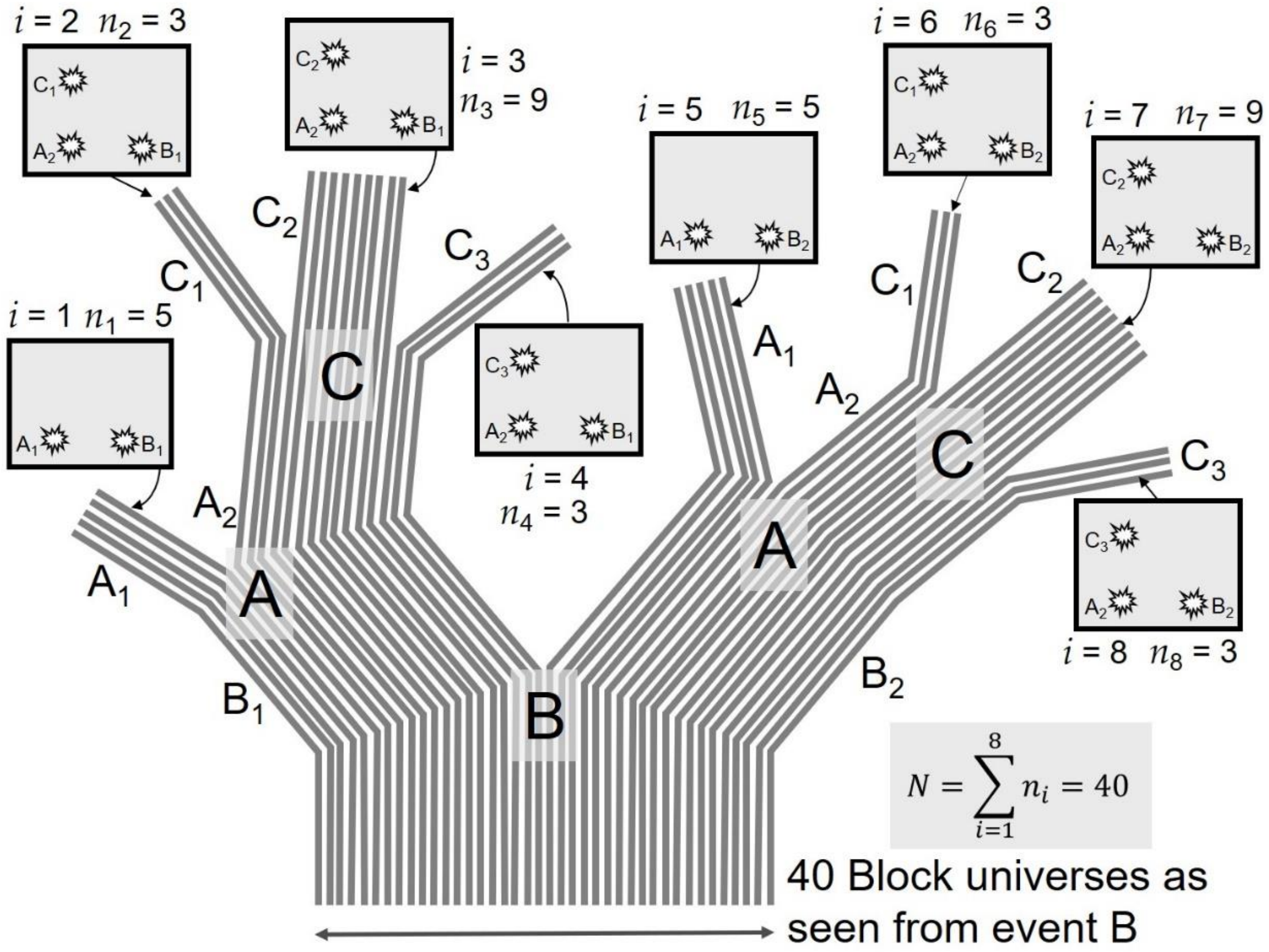

**Fig. 14**
This shows the Toy Multiverse from an alternative perspective to that in Fig. 13

Evidently, the nature of the pattern of the multiverse (namely, that the number of universes in the multiverse is discrete) is not compatible with the continuous Schrödinger equation. So, in our hypothesized multiverse, the Schrödinger equation would need to be rewritten as a digital equation (see, for instance, ’t Hooft (2014), Zahedi (2016)) The fact that no significant difference has been detected between measured and calculated quantum probabilities suggests that the number of parallel universes is large enough to mask any difference, although finding an appropriate experiment remains a possible test of the hypothesis (McKenzie 2016(a)).